\documentclass[final]{elsarticle}

\usepackage{times,amsmath,amssymb,subfigure,url,multirow,bm,color,booktabs}
\usepackage[flushleft]{threeparttable}
\DeclareMathOperator*{\argmin}{\arg\!\min}

\makeatletter
\newcommand{\tabincell}[2]{\begin{tabular}{@{}#1@{}}#2\end{tabular}}
\makeatother

\journal{Information Sciences}

\begin{document}

\begin{frontmatter}
\title{Transfer Learning for Motor Imagery Based Brain-Computer Interfaces: A Complete Pipeline}

\author[1,2]{Dongrui Wu}\ead{drwu@hust.edu.cn}
\author[1]{Xue Jiang}\ead{xuejiang@hust.edu.cn}
\author[1]{Ruimin Peng}\ead{rmpeng2019@hust.edu.cn}
\author[2]{Wanzeng Kong}\ead{kongwanzeng@hdu.edu.cn}
\author[1]{Jian Huang\corref{CA}}\cortext[CA]{Corresponding authors}\ead{huang\_jan@hust.edu.cn}
\author[1]{Zhigang Zeng\corref{CA}}\ead{zgzeng@hust.edu.cn}

\address[1]{Key Laboratory of the Ministry of Education for Image Processing and Intelligent Control,\\ School of Artificial Intelligence and Automation, \\ Huazhong University of Science and Technology, Wuhan 430074, China}
\address[2]{Zhejiang Key Laboratory for Brain-Machine Collaborative Intelligence,\\  Hangzhou Dianzi University, Hangzhou 310018, China}

\begin{abstract}
Transfer learning (TL) has been widely used in motor imagery (MI) based brain-computer interfaces (BCIs) to reduce the calibration effort for a new subject, and demonstrated promising performance. While a closed-loop MI-based BCI system, after electroencephalogram (EEG) signal acquisition and temporal filtering, includes spatial filtering, feature engineering, and classification blocks before sending out the control signal to an external device, previous approaches only considered TL in one or two such components. This paper proposes that TL could be considered in all three components (spatial filtering, feature engineering, and classification) of MI-based BCIs. Furthermore, it is also very important to specifically add a data alignment component before spatial filtering to make the data from different subjects more consistent, and hence to facilitate subsequential TL. Offline calibration experiments on two MI datasets verified our proposal. Especially, integrating data alignment and sophisticated TL approaches can significantly improve the classification performance, and hence greatly reduces the calibration effort.
\end{abstract}

\end{frontmatter}

\section{Introduction}

A brain-computer interface (BCI) \cite{Lance2012,Wolpaw2002} enables a user to communicate directly with an external device, e.g., a computer, using his/her brain signals, e.g., electroencephalogram (EEG). It can benefit both patients \cite{Pfurtscheller2008} and able-bodied people \cite{Erp2012,Nicolas-Alonso2012}.

Motor imagery (MI) \cite{Pfurtscheller2001} is a common paradigm in EEG-based BCIs, and also the focus of this paper. In MI-based BCIs, the user imagines the movements of his/her body parts, which activates different areas of the motor cortex of the brain, e.g., top-left for right-hand MI, top-right for left-hand MI, and top-central for feet MI. A classification algorithm can then be used to decode the recorded EEG signals and map the corresponding MI to a command for the external device.

The flowchart of a closed-loop EEG-based BCI system is shown in Figure~\ref{fig:BCI}. It consists of the following main components \cite{drwuTLBCI2021}:

\begin{figure}[htpb]\centering
\includegraphics[width=.6\linewidth,clip]{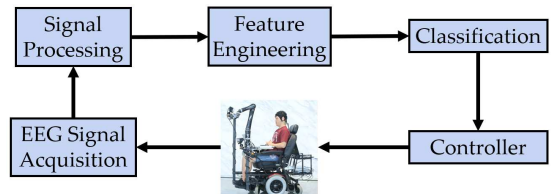}
\caption{A closed-loop EEG-based BCI system, using MI as an example. } \label{fig:BCI}
\end{figure}

\begin{enumerate}
\item \emph{Signal acquisition}, which uses a headset to collect EEG signal from the scalp, while the user is performing MI tasks.

\item \emph{Signal processing} \cite{Lotte2015}. Because EEG signals are weak, and easily contaminated by artifacts and interferences, e.g., muscle movements, eye blinks, heartbeats, powerline noise, etc., sophisticated signal processing approaches must be used to increase the signal-to-noise ratio. Both temporal filtering and spatial filtering are usually performed. Temporal filtering may include notch filtering to remove the 50Hz or 60Hz powerline interference, and then bandpass filtering, e.g., $[8,30]$ Hz, to remove DC drift and high frequency noise. Spatial filters \cite{drwuSF2018} include the basic ones, e.g., common average reference \cite{Teplan2002}, Laplacian filters \cite{Lagerlund1997},  principal component analysis \cite{Jolliffe2002}, etc., and more sophisticated ones, e.g., independent component analysis \cite{Delorme2004}, xDAWN \cite{Rivet2009}, canonical correlation analysis \cite{Roy2015}, common spatial patterns (CSP) \cite{Ramoser2000}, etc.

\item \emph{Feature engineering}, which includes feature extraction, and sometimes also feature selection. Time domain, frequency domain, time-frequency domain, Riemannian space \cite{Yger2017}, and/or topoplot features \cite{drwu4D2021} could be used.

\item \emph{Classification} \cite{Lotte2018}, which uses a machine learning algorithm to decode the EEG signal from the extracted features. Commonly used classifiers include linear discriminant analysis (LDA) and support vector machine (SVM).

\item \emph{Controller}, which sends a command to an external device, e.g., a wheelchair, according to the decoded EEG signal.
\end{enumerate}

Because of individual differences and non-stationarity of EEG signals, an MI-based BCI usually needs a long calibration session for a new subject, from 20-30 minutes \cite{Saha2018} to hours or even days. This lengthy calibration significantly reduces the utility of BCI systems. Hence, many sophisticated signal processing and machine learning approaches have been proposed recently to reduce or eliminate the calibration \cite{Dai2018,drwuLA2020,drwuEA2020,Jayaram2016,Rodrigues2019,Wang2015,drwuTHMS2017,drwuTLBCI2021,Zanini2018,drwuMEKT2020}.

One of the most promising such approaches is transfer learning (TL) \cite{Pan2010}, which uses data/knowledge from source domains (existing subjects) to help the calibration in the target domain (new subject). However, previous TL approaches for BCIs usually considered only one or two components of the closed-loop system in Figure~\ref{fig:BCI}, particularly, classification, as introduced in our latest survey \cite{drwuTLBCI2021}. For example, Jayaram \emph{et al.} \cite{Jayaram2016} proposed a multi-task learning (which is a subfield of TL) framework for cross-subject MI classification. To consider TL in spatial filtering, Dai \emph{et al.} \cite{Dai2018} proposed transfer kernel CSP to integrate kernel CSP \cite{Albalawi2012} and transfer kernel learning \cite{Long2015} for EEG trial filtering. To consider TL in feature engineering, Chen \emph{et al.} \cite{Chen2019} extended ReliefF \cite{Kononenko1994} and minimum redundancy maximum relevancy (mRMR) \cite{Peng2005} feature selection approaches to Class-Separate and Domain-Fused (CSDF)-ReliefF and CSDF-mRMR, which optimized both the class separability and the domain similarity simultaneously. They then further integrated CSDF-ReliefF and CSDF-mRMR with an adaptation regularization-based TL classifier \cite{Long2014}.

In this paper, we claim that TL should be considered in as many components of a BCI system as possible, and propose a complete TL pipeline for MI-based BCIs, shown in Figure~\ref{fig:TLBCI}:

\begin{figure}[htpb]\centering
\includegraphics[width=.7\linewidth,clip]{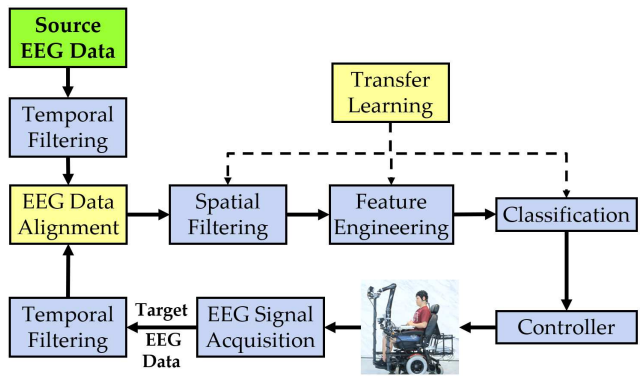}
\caption{A complete TL pipeline for closed-loop MI-based BCI systems.} \label{fig:TLBCI}
\end{figure}

\begin{enumerate}
\item \emph{Temporal filtering}, where band-pass filtering is performed on both the source and target domain data.

\item \emph{Data alignment}, which aligns EEG trials from the source domains and the target domain so that their distributions are more consistent. This is a new component, which does not exist in Figure~\ref{fig:BCI}, but will greatly facilitate TL in sequential components, as shown later in this paper.

\item \emph{Spatial filtering}, where TL can be used to design better spatial filters, especially when the amount of target domain labeled data is small.

\item \emph{Feature engineering}, where TL may be used to extract or select more informative features.

\item \emph{Classification}, where TL can be used to design better classifiers or regression models, especially when there are no or very few target domain labeled data.
\end{enumerate}

We will introduce some representative TL approaches in data alignment, spatial filtering, feature selection and classification, and demonstrate using two MI datasets that incorporate TL in all these components can indeed achieve better classification performance than not using TL, or using TL in only a subset of these components.

Our main contributions are:
\begin{enumerate}
\item We propose a complete TL pipeline for closed-loop MI-based BCI systems, as shown in Figure~\ref{fig:TLBCI}, and point out that explicitly including a data alignment component before spatial filtering is very important to the TL performance, for both traditional machine learning and deep learning, both offline and online classification, and both cross-subject and cross-session classification.

\item We verify through experiments that usually considering TL in more components in Figure~\ref{fig:TLBCI} can result in better classification performance, and more sophisticated TL approaches are usually more beneficial than simple TL approaches, or not using TL at all.
\end{enumerate}

The remainder of this paper is organized as follows: Section~\ref{sect:TL} introduces some representative TL approaches at different components of a BCI system. Section~\ref{sect:results} evaluates the performance of the complete TL pipeline in offline cross-subject MI classification. Section~\ref{sect:discussion} discusses the TL pipeline in offline cross-subject classification using deep learning, online cross-subject classification, and offline cross-session classification. Finally, Section~\ref{sect:conclusions} draws conclusions and points out some future research directions.

\section{TL Approaches} \label{sect:TL}

This section introduces the basic concepts of TL, and how TL could be used in data alignment, spatial filtering, feature selection and classification of an MI-based BCI system.

We consider offline binary classification only, and would like to use labeled EEG trials from a source subject to help classify trials from a target subject. When there are multiple source subjects, we can combine data from all source subjects and then view that as a single source subject, or perform TL for each source subject separately and then aggregate them.

Assume the source subject has $N_s$ labeled samples $\{X_s^n,y_s^n\}_{n=1}^{N_s}$, where $X_s^n\in\mathbb{R}^{c\times t}$ is the $n$-th EEG trial and $y_s^n$ the corresponding class label, in which $c$ is the number of EEG channels, and $t$ the number of time domain samples. The target subject has $N_l$ labeled samples $\{X_t^n,y_t^n\}_{n=1}^{N_l}$, and $N_u$ unlabeled samples $\{X_t^n\}_{n=N_l+1}^{N_l+N_u}$. In offline calibration, the $N_u$ unlabeled samples are also known to us, and we need to design a classifier to obtain their labels.

\subsection{TL}

TL \cite{Pan2010} uses data/knowledge from a source domain to help solve a task in a target domain. A domain consists of a feature space $\mathcal{X}$ and its associated marginal probability distribution $P(X)$, i.e., $\{\mathcal{X},P(X)\}$, where $X\in \mathcal{X}$. Two domains are different if they have different feature spaces, and/or different $P(X)$. A task consists of a label space $\mathcal{Y}$ and a prediction function $f(X)$, i.e., $\{\mathcal{Y},f(X)\}$. Two tasks are different if they have different label spaces, and/or different conditional probability distributions $P(y|X)$.

For BCI calibration, TL usually means to use labeled EEG trials from an existing subject to help the calibration for a new subject. This paper considers the scenario that both subjects have the same feature space and label space, i.e., the subjects wear the same EEG headset and perform the same types of MIs, but different $P(X)$ and $P(y|X)$. This is the most commonly encountered TL scenario in BCI calibration.

\begin{figure}[htpb]\centering
\includegraphics[width=.8\linewidth,clip]{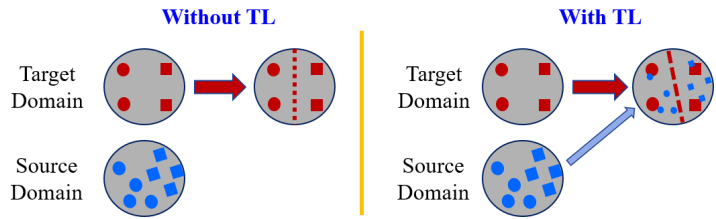}
\caption{Illustration of simple TL in classification. } \label{fig:TL}
\end{figure}

A very simple idea of TL in classifier training is illustrated in Figure~\ref{fig:TL}. Assume the target domain has only four training samples belonging to two classes (represented by different shapes), whereas the source domain has more. Without TL, we can build a classifier in the target domain using only its own four training samples. Since the number of training samples is very small, this classifier is usually unreliable. With TL, we can combine samples from the source domain with those in the target domain to train a classifier. Since the two domains may not be completely consistent, e.g., the marginal probability distributions may be different, we may assign the source domain samples smaller weights than the target domain samples. If optimized properly, the resulting classifier can usually achieve better generalization performance.

Figure~\ref{fig:TL} illustrates maybe the simplest TL approach in classification. Similar approaches may also be used in spatial filtering and feature engineering components in Figure~\ref{fig:TLBCI}. We will introduce some of them next.

\subsection{Euclidean Alignment (EA)} \label{sect:EA}

Due to individual differences, the marginal probability distributions of the EEG trials from different subjects are usually (significantly) different; so, it is very beneficial to perform data alignment to make different domains more consistent, before other operations in Figure~\ref{fig:TLBCI}.

\begin{figure}[htpb]\centering
\includegraphics[width=\linewidth,clip]{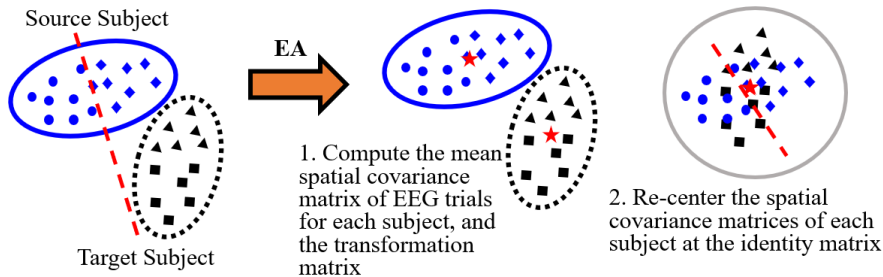}
\caption{EA for aligning EEG trials from different subjects (domains).} \label{fig:EA}
\end{figure}

Different EEG trial alignment approaches have been proposed recently \cite{drwuLA2020,drwuEA2020,Rodrigues2019,Zanini2018,drwuMEKT2020}. A summary and comparison of them is given in \cite{drwuTLBCI2021}. Among them, Euclidean alignment (EA), proposed by He and Wu \cite{drwuEA2020} and illustrated in Figure~\ref{fig:EA}, is easy to perform and completely unsupervised (does not need any labeled data from any subject). So, it is used as an example in this paper.

For the source subject, EA first computes
\begin{align}
  \bar{R}_s=\frac{1}{N_s}\sum_{n=1}^{N_s}X_s^n\left(X_s^n\right)^\top, \label{eq:ref2}
\end{align}
i.e., the Euclidean arithmetic mean of all spatial covariance matrices from the source subject, then performs the alignment by
\begin{align}
  \tilde{X}_s^n=\bar{R}_s^{-1/2}X_s^n. \label{eq:EA}
\end{align}
Similarly, for the target subject, EA computes the arithmetic mean of all $N_l+N_u$ spatial covariance matrices and then performs the alignment.

After EA, the aligned EEG trials are whitened \cite{drwuMEKT2020}, and their mean spatial covariance matrix from each subject equals the identity matrix \cite{drwuEA2020}; hence, the distributions of EEG trials from different subjects become more consistent. This will greatly benefit TL in subsequent steps.

\subsection{Pre-alignment Strategy (PS)}

Xu et al. \cite{Xu2020} proposed an online pre-alignment strategy (PS) to match the distributions from different domains. When used in offline classification, its formula is essentially identical to (\ref{eq:EA}), except that the Riemannian mean \cite{Pennec2006a} instead of the Euclidian mean is used in computing $\bar{R}_s$.

The \emph{Riemannian distance} $\delta(R,R^n)$ between two symmetric positive-definite covariance matrices $R\in\mathbb{R}^{c\times c}$ and $R^n\in\mathbb{R}^{c\times c}$ is the minimum length of a curve connecting them on the Riemannian manifold, computed as \cite{Moakher2005}:
\begin{align}
\delta(R,R^n)=\left\|\log \left(R^{-1}R^n\right)\right\|_F
=\left[\sum_{i=1}^c\log^2\lambda_i\right]^{\frac{1}{2}},
\end{align}
where the subscript $_F$ denotes the Frobenius norm, and $\lambda_i$, $i=1,...,c$, are the real eigenvalues of $R^{-1}R^n$.

The \emph{Riemannian mean} \cite{Pennec2006a} of $N$ covariance matrices is defined as the matrix minimizing the sum of the squared Riemannian distances, i.e.,
\begin{align}
\bar{R}=\arg\min\limits_R\sum_{n=1}^N\delta^2(R,R^n).
\end{align}
The Riemannian mean does not have a closed-form solution, but can be computed by an iterative gradient descent algorithm \cite{Fletcher2004}.

The characteristics of PS are almost identical to EA, i.e., it is completely unsupervised, and aligns the EEG trials directly, except that its computational cost is higher than EA, as the Riemannian mean does not have a closed-form solution.

\subsection{CSP} \label{sect:CSP}

CSP \cite{Blankertz2008,Ramoser2000} performs supervised spatial filtering for EEG trials, aiming to find a set of spatial filters to maximize the ratio of variance between two classes.

The traditional CSP uses data from the target subject only. For Class $k\in\{-1,1\}$, CSP tries to find a spatial filter matrix $W^{*}_k\in\mathbb{R}^{c\times f}$, where $f$ is the number of spatial filters, to maximize the variance ratio between Class $k$ and Class $-k$:
\begin{gather}
W^{*}_k=\arg\max_{W\in\mathbb{R}^{c\times f}}\dfrac{\mathrm{tr}(W^\top\bar{C}_t^kW)}{\mathrm{tr}(W^\top\bar{C}_t^{-k}W)}, \label{eq:Wk}
\end{gather}
where $\bar{C}_t^k\in\mathbb{R}^{c\times c}$ is the mean spatial covariance matrix of the $N_l$ labeled EEG trials in Class $k$, and $\mathrm{tr}$ the trace of a matrix. The solution $W^{*}_k$ is the concatenation of the $f$ leading eigenvectors of $(\bar{C}_t^{-k})^{-1}\bar{C}_t^k$.

Then, CSP concatenates the $2f$ spatial filters from both classes to obtain the complete filter matrix:
\begin{gather}
W^*=\begin{bmatrix}
W^{*}_{-1}&W^{*}_1
\end{bmatrix}\in\mathbb{R}^{c\times 2f}, \label{eq:W}
\end{gather}
and computes the spatially filtered $X_t^n$ by:
\begin{gather}
\tilde{X}_t^n=W^{*\top} X_t^n\in\mathbb{R}^{2f\times t}.
\end{gather}

Finally, the log-variances of $\tilde{X}_t^n$ can be extracted as features $\bm{x}_t^n\in\mathbb{R}^{1\times 2f}$ in later classification:
\begin{gather}
\bm{x}_t^n=\log\left( \dfrac{\mathrm{diag}\left(\tilde{X}_t^n\left(\tilde{X}_t^n\right)^\top\right)}
{\mathrm{tr}\left(\tilde{X}_t^n\left(\tilde{X}_t^n\right)^\top\right)} \right),\label{eq:CSP3}
\end{gather}
where $\mathrm{diag}$ means the diagonal elements of a matrix, and $\mathrm{log}$ is the logarithm operator.

\subsection{Combined CSP (CCSP)}

Because the target subject has very few labeled samples, i.e., $N_l$ is small, $W^*$ computed above may not be reliable. The source domain samples can be used to improve $W^*$.

In the combined CSP (CCSP), we simply concatenate the $N_s$ source domain labeled samples and $N_l$ target domain labeled samples to compute $W^*$. Note that all samples have the same weight, i.e., source domain and target domain samples are treated equally.

CCSP may be the simplest TL-based CSP approach.

\subsection{Regularized CSP (RCSP)}

Regularized CSP (RCSP) \cite{Lu2010} was specifically proposed to handle the problem that the target domain has very few labeled samples. Though the original paper did not mention TL, it actually used the idea of TL.

RCSP computes the regularized average spatial covariance matrix for Class~$k$ as:
\begin{align}
\hat{C}^k(\beta,\gamma)=(1-\gamma)\hat{C}^k(\beta)+\frac{\gamma}{c}\mathrm{tr}(\hat{C}^k(\beta))I,
\end{align}
where $\beta$ and $\gamma$ are two parameters in $[0,1]$, $I\in \mathbb{R}^{c\times c}$ is an identity matrix, and
\begin{align}
\hat{C}^k(\beta)=\frac{\beta N_l\bar{C}_t^k+(1-\beta) N_s\bar{C}_s^k}{\beta N_l+(1-\beta) N_s}.
\end{align}
$\hat{C}^k(\beta,\gamma)$ can then be used to replace $\bar{C}_t^k$ in (\ref{eq:Wk}) to compute the CSP filter matrix.

Note that when $\beta=1$ and $\gamma=0$, RCSP becomes the traditional CSP. When $\beta=0.5$ and $\gamma=0$, RCSP becomes CCSP.

\subsection{ReliefF}

ReliefF \cite{Kononenko1994} is a classical feature selection approach. Next we introduce its basic idea for  binary classification.

Let $x_i$ be the $i$-th feature, whose importance $w(x_i)$ is initialized to 0. ReliefF randomly selects a sample $\bm{x}$, and finds its $k$ ($k=10$ in this paper) nearest neighbors $H=\{\bm{h}_j\}_{j=1}^k$ in the same class, and also $k$ nearest neighbors $M=\{\bm{m}_j\}_{j=1}^k$ in the other class. It then updates $w(x_i)$ by
\begin{align}
w(x_i)&=w(x_i)-\frac{1}{k}\sum_{j=1}^k \mathrm{diff}(x_i,\bm{x},\bm{h}_j)
+\frac{1}{k}\sum_{j=1}^k \mathrm{diff}(x_i,\bm{x},\bm{m}_j), \label{eq:reliefF}
\end{align}
where $\mathrm{diff}(x_i,\bm{x},\bm{x}')$ denotes the difference between samples $\bm{x}$ and $\bm{x}'$ in terms of feature $x_i$. (\ref{eq:reliefF}) is very intuitive: the importance of $x_i$ should be decreased with its ability to discriminate samples from the same class [$\mathrm{diff}(x_i,\bm{x},\bm{h})$], and increased with its ability to discriminate samples from different classes [$\mathrm{diff}(x_i,\bm{x},\bm{m})$].

In this paper, ReliefF terminates after 100 iterations, i.e., 100 randomly selected $\bm{x}$ were used to compute the final $w(x_i)$. We then rank these $w(x_i)$ and select a few features corresponding to the largest $w(x_i)$.

\subsection{Combined ReliefF (CReliefF)}

The traditional ReliefF selects $\bm{x}$, $H$ and $M$ from only the target domain labeled samples. We propose a very simple TL extension of ReliefF, combined ReliefF (CReliefF), by selecting $\bm{x}$, $H$ and $M$ from labeled samples in both domains.

\subsection{LDA}

LDA is a popular linear classifier for binary classification. It assumes that the feature covariance matrices (not to be confused with the spatial covariance matrix of an EEG trial) from the two classes have full rank and are both equal to $\Sigma_t$. The classification for a new input $\bm{x}$ is then
\begin{align}
\mathrm{sign}\left(\bm{w}\bm{x}^\top-\theta\right),
\end{align}
where
\begin{align}
\bm{w}&=\Sigma_t^{-1}(\bar{\bm{x}}_{t,1}-\bar{\bm{x}}_{t,-1}),\\
\bm{\theta}&=\frac{1}{2}\bm{w}(\bar{\bm{x}}_{t,1}+\bar{\bm{x}}_{t,-1}),
\end{align}
in which $\bar{\bm{x}}_{t,-1}$ and $\bar{\bm{x}}_{t,1}$ are the mean feature vector of Class~$-1$ and Class~1 computed from the $N_l$ target domain labeled samples, respectively.

\subsection{Combined LDA (CLDA)}

When $N_l$ is small, the above LDA classifier may not be reliable. The combined LDA (CLDA) is a simple TL approach, which concatenates labeled samples from both the source domain and the target domain to train an LDA classifier. All samples from both domains are treated equally.

\subsection{Weighted Adaptation Regularization (wAR)}

Wu \cite{drwuTHMS2017} proposed weighted adaptation regularization (wAR), a TL approach for offline cross-subject EEG classification. Though the original experiments were conducted for event-related potential classification, wAR can also be used for MI classification.

wAR learns a classifier $f^*$ by minimizing the following regularized loss function:
\begin{align}
f^*=\argmin\limits_f&\sum_{n=1}^{N_s}w_s^n\ell(f(\bm{x}_s^n),y_s^n)
+w_t\sum_{n=1}^{N_l}w_t^n\ell(f(\bm{x}_t^n),y_t^n) \nonumber\\
& +\lambda_1\|f\|_K^2+\lambda_2 D_{f,K}(P_s(\bm{x}_s),P_t(\bm{x}_t))\nonumber \\
&+\lambda_3 D_{f,K}(P_s(\bm{x}_s|y_s),P_t(\bm{x}_t|y_t)) \label{eq:wAR}
\end{align}
where $\ell$ is the classification loss, $w_t$ is the overall weight of samples from the target subject, $w_s^n$ and $w_t^n$ are the weights for the $n$-th sample from the source subject and the target subject, respectively, $K$ is a kernel function, $P_s(\bm{x}_s)$ and $P_t(\bm{x}_t)$ are the marginal probability distribution of features from the source subject and the target subject, respectively, $P_s(\bm{x}_s|y_s)$ and $P_t(\bm{x}_t|y_t)$ are the conditional probability distribution from the source subject and the target subject, respectively, and $\lambda_1$, $\lambda_2$ and $\lambda_3$ are non-negative regularization parameters.

Briefly speaking, the five terms in (\ref{eq:wAR}) minimize the classification loss for the source subject, the classification loss for the target subject, the structural risk of the classifier, the distance between the marginal probability distributions of the two subjects, and the distance between the conditional probability distributions of the two subjects, respectively.

Although it looks complicated, (\ref{eq:wAR}) has a closed-form solution when the squared loss $\ell(f(\bm{x})-y)=(y-f(\bm{x}))^2$ is used \cite{drwuTHMS2017}.

\subsection{Online wAR (OwAR)}

Wu \cite{drwuTHMS2017} also proposed online wAR (OwAR), a TL approach for online cross-subject EEG classification.

OwAR learns a classifier $f^*$, also by minimizing the regularized loss function in (\ref{eq:wAR}). The only difference is that now the kernel matrix $K$ can only be computed from the $N_s$ labeled source domain samples and $N_l$ labeled target domain samples, but not the $N_u$ unlabeled target domain samples, which are unavailable in online classification.

\section{Experiments and Results} \label{sect:results}

This section evaluates the offline cross-subject classification performances of different combinations of TL approaches on two MI datasets.

\subsection{MI Datasets}

Two MI datasets from BCI Competition IV\footnote{http://www.bbci.de/competition/iv/.} were used in this study. They were also used in our previous research \cite{drwuLA2020,drwuEA2020,drwuMEKT2020}.

\begin{figure}[htpb]\centering
\includegraphics[width=\linewidth,clip]{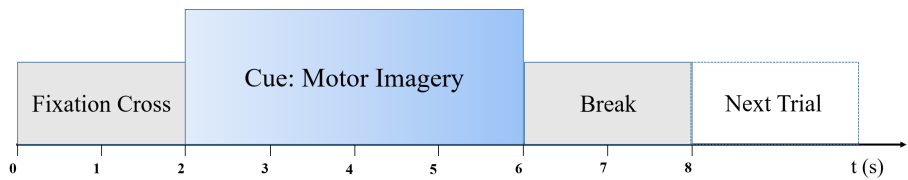}
\caption{Timing scheme of the MI tasks.} \label{fig:paradigm}
\end{figure}

In each experiment, the subject sat in front of a computer and performed visual cue based MI tasks, as shown in Figure~\ref{fig:paradigm}. A fixation cross on the black screen ($t=0$) prompted the subject to be prepared, and marked the start of a trial. After two seconds, a visual cue, which was an arrow pointing to a certain direction, was displayed for four seconds, during which the subject performed the instructed MI task. The  visual cue disappeared at $t=6$ second, and the MI also stopped. After a two-second break, the next trial started.

The first dataset\footnote{http://www.bbci.de/competition/iv/desc\_1.html.} (Dataset~1 \cite{Blankertz2007}) consisted of seven healthy subjects. Each subject performed two types of MIs, selected from three classes: left-hand, right-hand, and foot. We used the 59-channel EEG data collected from the calibration phase, which included complete marker information. Each subject had 100 trials per class.

The second MI dataset\footnote{http://www.bbci.de/competition/iv/desc\_2a.pdf.} (Dataset~2a) included nine heathy subjects. Each subject performed four different MIs: left-hand, right-hand, both feet, and tongue. We used the 22-channel EEG data and two classes of MIs (left-hand and right-hand) collected from the training phase. Each subject had 72 trials per class.

EEG data preprocessing steps were identical to those in \cite{drwuEA2020}. A causal [8, 30] Hz band-pass filter was used to remove muscle artifacts, powerline noise, and DC drift. Next, we extracted EEG signals between $[0.5, 3.5]$ seconds after the cue onset as our trials for both datasets.

\subsection{Algorithms}

We mainly compare the following 18 different algorithms, with various different configurations of TL components:
\begin{enumerate}
\item \emph{CSP-LDA}, which uses only the target domain labeled data to design the CSP filters, and then trains an LDA classifier on the target domain labeled data. No source data is used at all, i.e., no TL is used at all.
\item \emph{CSP-CLDA}, which uses only the target domain labeled data to design the CSP filters, and then trains a CLDA classifier by using labeled data from both domains, i.e., only the classifier uses a simple TL approach.
\item \emph{CSP-wAR}, which uses only the target domain labeled data to design the CSP filters, and then trains a wAR classifier by using data from both domains, i.e., only the classifier uses a sophisticated TL approach.
\item \emph{CCSP-LDA}, which concatenates labeled data from both domains to design the CCSP filters, and then trains an LDA classifier on target domain labeled data only, i.e., only spatial filtering uses a simple TL approach.
\item \emph{CCSP-CLDA}, which concatenates labeled data from both domains to design the CCSP filters, and then trains a CLDA classifier also from the concatenated data, i.e., both spatial filtering and classification use a simple TL approach.
\item \emph{CCSP-wAR}, which concatenates labeled data from both domains to design the CCSP filters, and then trains a wAR classifier also from the concatenated data, i.e., spatial filtering uses a simple TL approach, whereas classification uses a sophisticated TL approach.
\item \emph{RCSP-LDA}, which uses labeled data from both domains to design the RCSP filters, and then trains an LDA classifier from the target domain labeled data only, i.e., spatial filtering uses a sophisticated TL approach, whereas classification does not use TL at all.
\item \emph{RCSP-CLDA}, which uses labeled data from both domains to design the RCSP filters, and then trains a CLDA classifier also from this concatenated data, i.e., spatial filtering uses a sophisticated TL approach, whereas classification uses a simple TL approach.
\item \emph{RCSP-wAR}, which uses labeled data from both domains to design the RCSP filters, and then trains a wAR classifier from this concatenated data, i.e., both spatial filtering and classification use a sophisticated TL approach.
\item \emph{EA-CSP-LDA}, which performs EA before CSP-LDA.
\item \emph{EA-CSP-CLDA}, which performs EA before CSP-CLDA, i.e., only the classifier uses a simple TL approach, after EA.
\item \emph{EA-CSP-wAR}, which performs EA before CSP-wAR, i.e., only the classifier uses a sophisticated TL approach, after EA.
\item \emph{EA-CCSP-LDA}, which performs EA before CCSP-LDA, i.e., only spatial filtering uses a simple TL approach, after EA.
\item \emph{EA-CCSP-CLDA}, which performs EA before CCSP-CLDA, i.e., both spatial filtering and classification use a simple TL approach, after EA.
\item \emph{EA-CCSP-wAR}, which performs EA before CCSP-wAR, i.e., spatial filtering uses a simple TL approach, whereas classification uses a sophisticated TL approach, after EA.
\item \emph{EA-RCSP-LDA}, which performs EA before RCSP-LDA, i.e., only spatial filtering uses a sophisticated TL approach, after EA.
\item \emph{EA-RCSP-CLDA}, which performs EA before RCSP-CLDA, i.e., spatial filtering uses a sophisticated TL approach, whereas classification uses a simple TL approach, after EA.
\item \emph{EA-RCSP-wAR}, which performs EA before RCSP-wAR, i.e., both spatial filtering and classification use a sophisticated TL approach, after EA.
\end{enumerate}
Additionally, there were nine PS based approaches, which replace EA in the nine EA based approaches by PS, respectively. A summary of the 27 algorithms is shown in Table~\ref{tab:0}.

\begin{table}[htpb]
\centering \renewcommand{\arraystretch}{1.2} \setlength{\tabcolsep}{1.5mm}
\caption{Summary of the 27 algorithms with various degrees of TL.}  \label{tab:0}
\begin{tabular}{c|c|cc|cc} \toprule
& Data  & \multicolumn{2}{|c|}{Spatia Filtering} & \multicolumn{2}{c}{Classifier} \\  \cline{3-6}
Algorithm& Alignment& Simple TL & Sophisticated TL & Simple TL & Sophisticated TL \\ \midrule
CSP-LDA &     --    & --  & --  &-- &  --    \\
CSP-CLDA&     --  &--& --   &\checkmark& -- \\
CSP-wAR&       --  & --  & -- & -- &\checkmark    \\
CCSP-LDA&      --  & \checkmark  &-- & -- & --     \\
CCSP-CLDA&     -- &   \checkmark&--&\checkmark&   --   \\
CCSP-wAR&     --  &   \checkmark&--&--&\checkmark   \\
RCSP-LDA&     --  &--   &  \checkmark &--& -- \\
RCSP-CLDA&   --   & --  & \checkmark&\checkmark& --   \\
RCSP-wAR&     --  & --  &  \checkmark&--&\checkmark   \\ \midrule
EA-CSP-LDA&    \checkmark &  -- &--    &--& --\\
EA-CSP-CLDA&   \checkmark &--   &-- &\checkmark&--    \\
EA-CSP-wAR&    \checkmark & --  &-- &--&\checkmark    \\
EA-CCSP-LDA&   \checkmark &  \checkmark &  --  &--& --\\
EA-CCSP-CLDA&  \checkmark &   \checkmark&--  &\checkmark&--   \\
EA-CCSP-wAR&   \checkmark &   \checkmark&-- &--&\checkmark    \\
EA-RCSP-LDA&   \checkmark &  -- & \checkmark&--&--    \\
EA-RCSP-CLDA&  \checkmark & --  & \checkmark&\checkmark&  --  \\
EA-RCSP-wAR&   \checkmark & --  &  \checkmark&--&\checkmark   \\ \midrule
PS-CSP-LDA&    \checkmark &  -- &--    &--& --\\
PS-CSP-CLDA&   \checkmark & --  &-- &\checkmark&  --  \\
PS-CSP-wAR&    \checkmark &  -- & --&--&\checkmark    \\
PS-CCSP-LDA&   \checkmark &  \checkmark & --   &--&-- \\
PS-CCSP-CLDA&  \checkmark &   \checkmark& -- &\checkmark& --  \\
PS-CCSP-wAR&   \checkmark &   \checkmark&-- &--&\checkmark    \\
PS-RCSP-LDA&   \checkmark & --  & \checkmark&--&--    \\
PS-RCSP-CLDA&  \checkmark &  -- & \checkmark&\checkmark&    \\
PS-RCSP-wAR&   \checkmark &  -- &  \checkmark&--&\checkmark   \\  \bottomrule
\end{tabular}
\end{table}

Six (a typical number \cite{Rao2010}) spatial filters were used in all CSP algorithms. $\beta=\gamma=0.1$ were used in RCSP. $w_t=10$, $\lambda_1=0.1$, $\lambda_2=\lambda_3=10$ were used in wAR, as in \cite{drwuTHMS2017,drwuTNSRE2016}, except that $w_t$ was increased from 2 to 10 because the combined source domain has much more labeled samples than the target domain. The source code is available at https://github.com/drwuHUST/TLBCI.

By comparing between different pairs of the above algorithms, we can individually study the effect of TL in different components of Figure~\ref{fig:TLBCI}.

\subsection{Experimental Settings and Results}

For each dataset, we sequentially selected one subject as the target subject and all remaining ones as the source subjects, i.e., we performed cross-subject evaluations. As in \cite{drwuEA2020}, we combined all source subjects as a single source domain, and performed the corresponding TL. This procedure was repeated for each subject, so that each one became the target subject once.

The number of randomly selected labeled samples in the target domain ($N_l$) increased from zero to 20, with a step of 4. Because there was randomness involved, we repeated this process 30 times and report the average results.

Note that for algorithms whose spatial filtering component did not involve TL, e.g., those with \emph{CSP-}, when $N_l=0$, no CSP filters can be trained, and hence no model can be built. All other algorithms used TL in CSP, and hence the source domain labeled data can be used to train the CSP filters even when $N_l=0$. Similarly, for algorithms whose classifier did not involve TL, e.g., those with $\emph{-LDA}$, when $N_l=0$, no LDA classifier can be trained.

Note also that since we consider offline classification, all unlabeled samples in the target domain are known, and can be used in EA, PS and wAR. There is no data leakage here.

\begin{figure*}[htpb] \centering
\subfigure[]{\label{fig:MI1}     \includegraphics[width=\linewidth,clip]{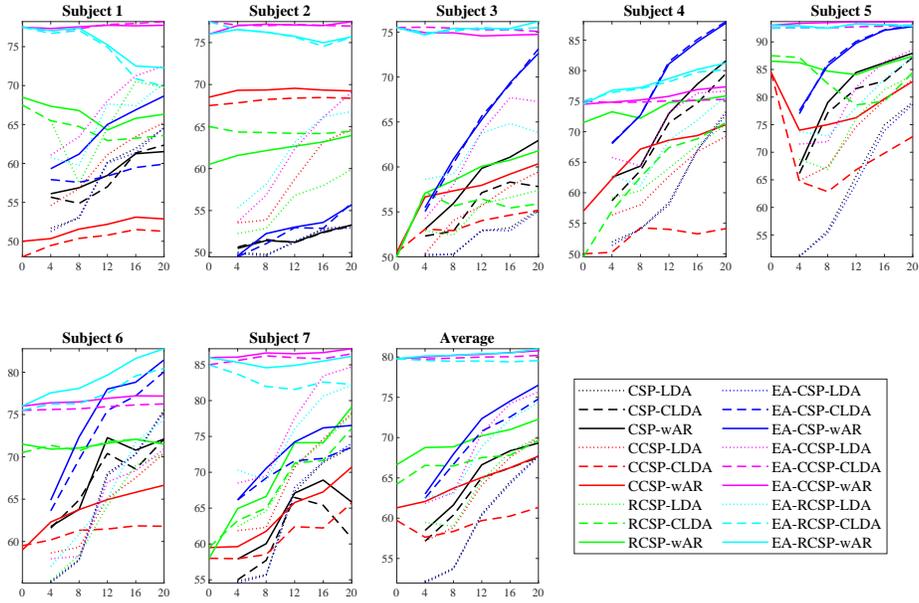}}
\subfigure[]{\label{fig:MI2}     \includegraphics[width=\linewidth,clip]{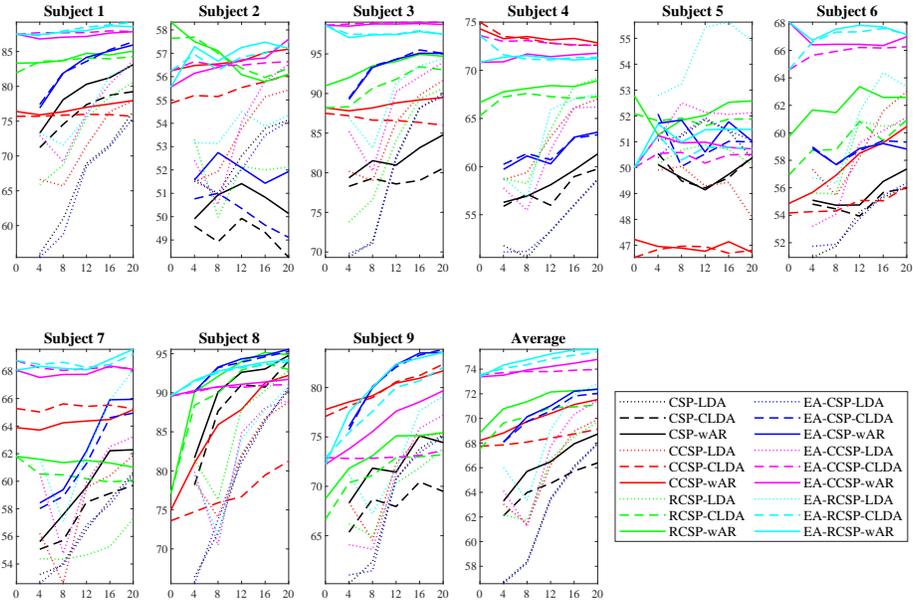}}
\caption{Offline cross-subject classification accuracies (vertical axis) on the MI datasets, with different $N_l$ (horizontal axis). (a) Dataset~1; (b) Dataset~2a.} \label{fig:MI}
\end{figure*}

The cross-subject classification accuracies, averaged over 30 random runs, are shown in Figure~\ref{fig:MI}. The average performances over all subjects are shown in the last panel of each subfigure. To ensure that the curves are distinguishable, we omitted the curves from the nine PS based algorithms, which were very similar to their EA counterparts.

\subsection{The General Effect of TL}

In Figure~\ref{fig:MI}, by comparing CSP-LDA, which did not use TL at all, with the other 17 algorithms, which used simple or sophisticated TL in one or more components of Figure~\ref{fig:TLBCI}, we can see that when $N_l$ was small, TL almost always resulted in better performance, no matter how much TL was used. However, when $N_l$ increased, CSP-LDA gradually outperformed certain simple TL approaches, e.g., CSP-CLDA and CCSP-CLDA, whereas sophisticated TL approaches, e.g., EA-RCSP-wAR, almost always outperformed CSP-LDA. These results suggest that sophisticated TL may always be beneficial.

To quantitatively study the general effect of TL, we computed the mean classification accuracies of the 27 approaches when $N_l$ increased from 4 to 20 (we did not use $N_l=0$ because certain approaches did not work in this case), and compared them with that of CSP-LDA. The results are shown in Table~\ref{tab:1}. We also performed paired $t$-tests to verify if the performance improvements over CSP-LDA were statistically significant ($\alpha=0.05$), and marked the insignificant ones by an underline. Table~\ref{tab:1} confirms again that generally more sophisticated TL approaches achieved larger performance improvements.


\begin{table}[htpb]
\centering \renewcommand{\arraystretch}{1.2} \setlength{\tabcolsep}{1.5mm}
\caption{Offline cross-subject classification accuracies (mean$\pm$std) of different TL approaches, and their improvements over CSP-LDA. Performance improvements not statistically significant ($\alpha=0.05$) are marked with an underline.}
 \label{tab:1}
\begin{tabular}{c|cc|cc} \toprule
\multirow{3}{*}{Algorithm}&\multicolumn{2}{c|}{Dataset 1} &\multicolumn{2}{|c}{Dataset 2a}\\ \cline{2-5}
& Accuracy & Improvement & Accuracy & Improvement \\
& (\%) & (\%) & (\%) & (\%) \\ \midrule
CSP-LDA &    59.75$\pm$6.74&     -- &   62.54$\pm$4.88&     --    \\
CSP-CLDA&    63.27$\pm$4.37&    5.91&   64.59$\pm$1.67&    3.27   \\
CSP-wAR&     64.88$\pm$4.65&    8.60&   66.43$\pm$2.12&    6.22   \\
CCSP-LDA&    63.54$\pm$5.52&    6.35&   65.91$\pm$3.67&    5.39   \\
CCSP-CLDA&   59.45$\pm$1.48&   \underline{-0.49}&   68.44$\pm$0.51&    9.43   \\
CCSP-wAR&    64.95$\pm$2.18&    8.72&   70.32$\pm$1.08&   12.44   \\
RCSP-LDA&    64.28$\pm$5.01&    7.59&   65.76$\pm$3.65&    5.15   \\
RCSP-CLDA&   67.60$\pm$1.29&   13.15&   70.54$\pm$0.63&   12.78   \\
RCSP-wAR&    70.22$\pm$1.49&   17.54&   71.78$\pm$0.68&   14.77   \\ \midrule
EA-CSP-LDA&  59.57$\pm$6.72&   -0.29&   62.43$\pm$4.89&   \underline{-0.18}   \\
EA-CSP-CLDA& 69.46$\pm$4.91&   16.27&   70.46$\pm$1.62&   12.66   \\
EA-CSP-wAR&  70.87$\pm$5.43&   18.61&   70.76$\pm$1.76&   13.14   \\
EA-CCSP-LDA& 69.09$\pm$6.32&   15.64&   66.97$\pm$4.28&    7.08   \\
EA-CCSP-CLDA&79.97$\pm$0.17&   33.85&   73.84$\pm$0.10&   18.07   \\
EA-CCSP-wAR& 80.37$\pm$0.30&   34.52&   74.19$\pm$0.50&   18.62   \\
EA-RCSP-LDA& 68.24$\pm$5.31&   14.21&   68.64$\pm$3.99&    9.75   \\
EA-RCSP-CLDA&79.51$\pm$0.08&   33.08&   74.76$\pm$0.55&   19.54   \\
EA-RCSP-wAR& 80.45$\pm$0.35&   34.66&   75.10$\pm$0.54&   20.08   \\ \midrule
PS-CSP-LDA&  59.58$\pm$6.73&   -0.28&   62.47$\pm$4.90&   \underline{-0.12}   \\
PS-CSP-CLDA& 71.51$\pm$5.26&   19.69&   70.49$\pm$1.59&   12.71   \\
PS-CSP-wAR&  71.89$\pm$5.17&   20.33&   70.85$\pm$1.70&   13.29   \\
PS-CCSP-LDA& 67.59$\pm$5.72&   13.13&   66.48$\pm$4.17&    6.29   \\
PS-CCSP-CLDA&78.35$\pm$0.18&   31.14&   73.87$\pm$0.14&   18.11   \\
PS-CCSP-wAR& 79.15$\pm$0.14&   32.48&   74.67$\pm$0.47&   19.39   \\
PS-RCSP-LDA& 68.92$\pm$5.25&   15.36&   68.02$\pm$3.89&    8.77   \\
PS-RCSP-CLDA&79.15$\pm$0.13&   32.48&   74.22$\pm$0.66&   18.68   \\
PS-RCSP-wAR& 79.56$\pm$0.17&   33.17&   74.68$\pm$0.70&   19.40   \\ \bottomrule
\end{tabular}
\end{table}

\subsection{The Effect of Data Alignment}

In Figure~\ref{fig:MI}, comparing algorithms without EA and their counterparts with EA, e.g., CSP-CLDA and EA-CSP-CLDA, we can observe that every EA version almost always significantly outperformed its non-EA counterpart. Similar observations can also be made for PS. These results suggest that a data alignment approach
such as EA or PS should always be included as a TL preprocessing step in a BCI system.

To quantitatively verify the above conclusion, we also show the mean classification accuracies of algorithms without and with EA/PS in Table~\ref{tab:2}. Clearly, EA and PS significantly improved the classification accuracies when TL is used in at least one component of spatial filtering and classification, especially on Dataset~1.

Interestingly, when CSP-LDA was used, i.e., no TL was used at all in spatial filtering and classification, EA/PS slightly reduced the classification performance. We were not able to find an explanation for this; however, when no TL will be performed, there is no point to align EEG trials from different subjects. So, this negative transfer will not happen in practice, and should not be a concern.

\begin{table}[htpb]
\centering \renewcommand{\arraystretch}{1.2} \setlength{\tabcolsep}{1.2mm} \small
\caption{Offline cross-subject classification accuracies (mean$\pm$std) of algorithms without and with EA/PS, and the improvements of the latter over the former. Performance improvements not statistically significant ($\alpha=0.05$) are marked with an underline.} \label{tab:2}
\begin{tabular}{c|c|ccc|ccc} \toprule
   &  &w/o EA &w/ EA  &Imp. &w/o PS  &w/ PS  &Imp. \\
Dataset&Algorithm& (\%) & (\%) & (\%)& (\%) & (\%) &(\%)\\\midrule
\multirow{8}{*}{1}
&CSP-LDA &      59.75$\pm$6.74&   59.57$\pm$6.72&   -0.29 &  59.75$\pm$6.74&   59.58$\pm$6.73&   -0.28\\
&CSP-CLDA&      63.27$\pm$4.37&   69.46$\pm$4.91&    9.79 &  63.27$\pm$4.37&   71.51$\pm$5.26&   13.02\\
&CSP-wAR&       64.88$\pm$4.65&   70.87$\pm$5.43&    9.22 &  64.88$\pm$4.65&   71.89$\pm$5.17&   10.80\\
&CCSP-LDA&      63.54$\pm$5.52&   69.09$\pm$6.32&    8.74 &  63.54$\pm$5.52&   67.59$\pm$5.72&    6.37\\
&CCSP-CLDA&     59.45$\pm$1.48&   79.97$\pm$0.17&   34.51 &  59.45$\pm$1.48&   78.35$\pm$0.18&   31.78\\
&CCSP-wAR&      64.95$\pm$2.18&   80.37$\pm$0.30&   23.74 &  64.95$\pm$2.18&   79.15$\pm$0.14&   21.86\\
&RCSP-LDA&      64.28$\pm$5.01&   68.24$\pm$5.31&    6.16 &  64.28$\pm$5.01&   68.92$\pm$5.25&    7.22\\
&RCSP-CLDA&     67.60$\pm$1.29&   79.51$\pm$0.08&   17.61 &  67.60$\pm$1.29&   79.15$\pm$0.13&   17.08\\
&RCSP-wAR&      70.22$\pm$1.49&   80.45$\pm$0.35&   14.56 &  70.22$\pm$1.49&   79.56$\pm$0.17&   13.29\\ \midrule \multirow{8}{*}{2a}
&CSP-LDA&       62.54$\pm$4.88&   62.43$\pm$4.89&   \underline{-0.18} &  62.54$\pm$4.88&   62.47$\pm$4.90&   \underline{-0.12}\\
&CSP-CLDA&      64.59$\pm$1.67&   70.46$\pm$1.62&    9.09 &  64.59$\pm$1.67&   70.49$\pm$1.59&    9.14\\
&CSP-wAR&       66.43$\pm$2.12&   70.76$\pm$1.76&    6.52 &  66.43$\pm$2.12&   70.85$\pm$1.70&    6.66\\
&CCSP-LDA&      65.91$\pm$3.67&   66.97$\pm$4.28&    1.60 &  65.91$\pm$3.67&   66.48$\pm$4.17&    \underline{0.86}\\
&CCSP-CLDA&     68.44$\pm$0.51&   73.84$\pm$0.10&    7.89 &  68.44$\pm$0.51&   73.87$\pm$0.14&    7.93\\
&CCSP-wAR&      70.32$\pm$1.08&   74.19$\pm$0.50&    5.49 &  70.32$\pm$1.08&   74.67$\pm$0.47&    6.18\\
&RCSP-LDA&      65.76$\pm$3.65&   68.64$\pm$3.99&    4.38 &  65.76$\pm$3.65&   68.02$\pm$3.89&    3.44\\
&RCSP-CLDA&     70.54$\pm$0.63&   74.76$\pm$0.55&    5.99 &  70.54$\pm$0.63&   74.22$\pm$0.66&    5.22\\
&RCSP-wAR&      71.78$\pm$0.68&   75.10$\pm$0.54&    4.62 &  71.78$\pm$0.68&   74.68$\pm$0.70&    4.03 \\ \bottomrule
\end{tabular}
\end{table}     \normalsize

\subsection{The Effect of TL in Spatial Filtering}

In Figure~\ref{fig:MI}, comparing algorithms without TL in spatial filtering (CSP), with simple TL in spatial filtering (CCSP), and with sophisticated TL in spatial filtering (RCSP), e.g., CSP-CLDA, CCSP-CLDA and RCSP-CLDA, we can observe that simple TL in spatial filtering may not always work (e.g., CCSP-CLDA had worse performance than CSP-CLDA on Dataset~1, but better performance on Dataset~2a), but sophisticated TL in spatial filtering was almost always beneficial (e.g., RCSP-CLDA outperformed both CSP-CLDA and CCSP-CLDA on both datasets). So, sophisticated TL approaches, such as RCSP, should be used in spatial filtering in a BCI system.

To quantitatively verify the above conclusion, we also show the mean classification accuracies of algorithms without and with TL in spatial filtering in Table~\ref{tab:3}. Clearly, RCSP (sophisticated TL in spatial filtering) always outperformed the corresponding CSP (no TL in spatial filtering) and CCSP (simple TL in spatial filtering) versions.

\begin{table}[htpb]
\centering \renewcommand{\arraystretch}{1.2} \setlength{\tabcolsep}{1.5mm}
\caption{The effect of TL in spatial filtering. Performance improvements not statistically significant ($\alpha=0.05$) are marked with an underline.} \label{tab:3}
\begin{tabular}{c|c|c|cc|cc} \toprule
\multirow{3}{*}{Dataset}&\multirow{3}{*}{Algorithm}& No TL
& \multicolumn{2}{|c|}{Simple TL} & \multicolumn{2}{|c}{Sophisticated TL} \\ \cline{3-7}
& &\tabincell{c}{CSP}&\tabincell{c}{CCSP}&\tabincell{c}{Imp. \\ (\%)} &\tabincell{c}{RCSP}&\tabincell{c}{Imp. \\ (\%)}\\ \midrule \multirow{9}{*}{1}
&CSP-LDA&      59.75$\pm$6.74&   63.54$\pm$5.52&    6.35&   64.28$\pm$5.01&    7.59\\
&CSP-CLDA&     63.27$\pm$4.37&   59.45$\pm$1.48&   -6.04&   67.60$\pm$1.29&    6.84\\
&CSP-wAR&      64.88$\pm$4.65&   64.95$\pm$2.18&    \underline{0.11}&   70.22$\pm$1.49&    8.23\\
& EA-CSP-LDA & 59.57$\pm$6.72&   69.09$\pm$6.32&   15.98&   68.24$\pm$5.31&   14.55\\
& EA-CSP-CLDA &69.46$\pm$4.91&   79.97$\pm$0.17&   15.12&   79.51$\pm$0.08&   14.46\\
& EA-CSP-wAR  &70.87$\pm$5.43&   80.37$\pm$0.30&   13.41&   80.45$\pm$0.35&   13.53\\
& PS-CSP-LDA & 59.58$\pm$6.73&   67.59$\pm$5.72&   13.44&   68.92$\pm$5.25&   15.68\\
& PS-CSP-CLDA &71.51$\pm$5.26&   78.35$\pm$0.18&    9.56&   79.15$\pm$0.13&   10.68\\
& PS-CSP-wAR  &71.89$\pm$5.17&   79.15$\pm$0.14&   10.10&   79.56$\pm$0.17&   10.66\\ \midrule \multirow{9}{*}{2a}
&CSP-LDA&      62.54$\pm$4.88&   65.91$\pm$3.67&    5.39&   65.76$\pm$3.65&    5.15\\
&CSP-CLDA&     64.59$\pm$1.67&   68.44$\pm$0.51&    5.96&   70.54$\pm$0.63&    9.21\\
&CSP-wAR&      66.43$\pm$2.12&   70.32$\pm$1.08&    5.86&   71.78$\pm$0.68&    8.06\\
& EA-CSP-LDA  &62.43$\pm$4.89&   66.97$\pm$4.28&    7.28&   68.64$\pm$3.99&    9.95\\
& EA-CSP-CLDA &70.46$\pm$1.62&   73.84$\pm$0.10&    4.80&   74.76$\pm$0.55&    6.11\\
& EA-CSP-wAR  &70.76$\pm$1.76&   74.19$\pm$0.50&    4.84&   75.10$\pm$0.54&    6.13\\
& PS-CSP-LDA & 62.47$\pm$4.90&   66.48$\pm$4.17&    6.42&   68.02$\pm$3.89&    8.90\\
& PS-CSP-CLDA &70.49$\pm$1.59&   73.87$\pm$0.14&    4.79&   74.22$\pm$0.66&    5.29\\
& PS-CSP-wAR  &70.85$\pm$1.70&   74.67$\pm$0.47&    5.39&   74.68$\pm$0.70&    5.40\\ \bottomrule
\end{tabular}
\end{table}

\subsection{The Effect of TL in Feature Selection}

Assume $2f$ spatial filters are needed. Then, as introduced in Section~\ref{sect:CSP}, in traditional CSP, $f$ of them are the leading eigenvectors of $(\bar{C}^N)^{-1}\bar{C}^P$, and the other $f$ are the leading eigenvectors of $(\bar{C}^P)^{-1}\bar{C}^N$, where $\bar{C}^P$ and $\bar{C}^N$ are the mean covariance matrices of the positive and negative classes, respectively. $f=3$ is typically used in the literature.

In this subsection, in order to show the effect of TL in feature selection, we use $f=10$ to extract 20 CSP filters, compute the 20 corresponding features in (\ref{eq:CSP3}), and then use different versions of ReliefF to select the best six among them. More specifically, without data alignment, we compared the following algorithms:
\begin{enumerate}
\item \emph{CSP6-LDA}, which is identical to \emph{CSP-LDA} in previous subsections. Here we add `6' to emphasize that it uses six CSP filters, obtained from the leading eigenvectors.
\item \emph{CSP20-ReliefF6-LDA}, which uses the traditional CSP algorithm to extract 20 spatial filters, compute the 20 corresponding features in (\ref{eq:CSP3}), and then use ReliefF to select the best six from them. ReliefF uses the target domain labeled data only, i.e., no TL is used.
\item \emph{CSP20-CReliefF6-LDA}, which uses the traditional CSP to extract 20 filters, compute the 20 corresponding features in (\ref{eq:CSP3}), and then use CReliefF to select the best six from them. CReliefF uses labeled data from both domains, so there is TL.
\end{enumerate}
LDA above can also be replaced by wAR, and CSP by RCSP. So, there could be 12 different configurations. Additionally, EA can also be added before each algorithm. So, there are a total of 24 different algorithms.

The cross-subject classification results of the 24 algorithms are shown in Table~\ref{tab:4}. When CReliefF was used to select the best six spatial filters from the 20 candidates (CSP20$\rightarrow$CReliefF6), the resulting classification performance was generally better than using ReliefF directly (CSP20$\rightarrow$ReliefF6), i.e., TL could be helpful in feature selection. However, using the six leading eigenvectors in CSP (CSP6) generally gave the best performance, justifying the common practice in the literature.

In summary, we have shown that if feature selection must be performed, then using TL may improve the performance; however, when CSP filters are used, using the leading eigenvectors is good enough, and we can safely omit feature selection.

\begin{table}[htpb]
\centering \renewcommand{\arraystretch}{1.2} \setlength{\tabcolsep}{1.3mm} \small
\caption{The effect of TL in feature selection. Performance improvements not statistically significant ($\alpha=0.05$) are marked with an underline.} \label{tab:4}
\begin{tabular}{c|c|c|c|cc} \toprule
\multirow{3}{*}{Dataset}&\multirow{3}{*}{Algorithm}& \multicolumn{2}{|c}{No TL}
 & \multicolumn{2}{|c}{TL} \\ \cline{3-6}
& &CSP6  &\tabincell{c}{CSP20$\rightarrow$ReliefF6} &\tabincell{c}{CSP20$\rightarrow$CReliefF6}&\tabincell{c}{Imp. (\%)} \\  \midrule \multirow{8}{*}{1}
&CSP-LDA&      59.91$\pm$6.72&   58.91$\pm$6.47&   59.74$\pm$6.23&    1.40\\
&CSP-wAR&      64.21$\pm$5.14&   61.20$\pm$5.54&   63.86$\pm$5.22&    4.36\\
&RCSP-LDA&     64.74$\pm$5.36&   61.38$\pm$5.97&   63.89$\pm$5.50&    4.09\\
& RCSP-wAR &   70.85$\pm$1.08&   64.67$\pm$3.55&   70.71$\pm$0.96&    9.35\\
& EA-CSP-LDA & 59.79$\pm$6.61&   58.02$\pm$5.79&   60.75$\pm$6.69&    4.71\\
& EA-CSP-wAR & 70.48$\pm$5.66&   65.68$\pm$5.93&   71.26$\pm$6.35&    8.50\\
& EA-RCSP-LDA &69.52$\pm$5.25&   63.51$\pm$5.89&   68.11$\pm$6.10&    7.24\\
& EA-RCSP-wAR &81.63$\pm$0.91&   72.78$\pm$3.88&   81.45$\pm$1.01&   11.92\\\midrule \multirow{8}{*}{2a}
&CSP-LDA&      62.32$\pm$5.63&   61.86$\pm$5.51&   62.82$\pm$4.98&    1.55\\
&CSP-wAR&      66.21$\pm$2.50&   64.72$\pm$3.12&   65.29$\pm$2.49&    0.89\\
&RCSP-LDA&     65.35$\pm$4.11&   62.97$\pm$5.12&   64.78$\pm$4.17&    2.88\\
& RCSP-wAR &   71.58$\pm$0.73&   68.08$\pm$2.40&   70.79$\pm$0.98&    3.98\\
& EA-CSP-LDA & 62.28$\pm$5.66&   61.24$\pm$5.70&   61.43$\pm$4.60&    \underline{0.31}\\
& EA-CSP-wAR & 70.50$\pm$2.38&   69.11$\pm$3.11&   68.13$\pm$1.67&   -1.42\\
& EA-RCSP-LDA &68.50$\pm$4.13&   64.19$\pm$4.68&   62.37$\pm$3.94&   -2.83\\
& EA-RCSP-wAR &75.00$\pm$0.59&   72.42$\pm$1.63&   69.55$\pm$0.38&   -3.96\\  \bottomrule
\end{tabular}
\end{table} \normalsize

\subsection{The Effect of TL in the Classifier}

In Figure~\ref{fig:MI}, comparing algorithms with simple and sophisticated TL in the classifier, e.g., CCSP-CLDA and CCSP-wAR, we can observe that sophisticated TL in the classifier almost always outperformed simple TL, regardless of whether TL was used in other components or not. So, sophisticated TL approaches, such as wAR, should be used in the classifier in a BCI system.

To quantitatively verify the above conclusion, we also show the mean classification accuracies of algorithms without and with TL in the classifier in Table~\ref{tab:5}. Clearly, on average wAR (sophisticated TL in the classifier) always outperformed CLDA (simple TL in the classifier).

Interestingly, when EA or PS was used, the performance improvement of wAR over CLDA became smaller, because EA or PS reduced the discrepancy between the source and target domain data, and hence made classification easier.

\begin{table}[htpb]
\centering \renewcommand{\arraystretch}{1.2} \setlength{\tabcolsep}{1mm}
\caption{The effect of TL in the classifier. Performance improvements not statistically significant ($\alpha=0.05$) are marked with an underline.} \label{tab:5}
\begin{tabular}{c|c|c|cc|cc} \toprule
\multirow{3}{*}{Dataset}&\multirow{3}{*}{Algorithm}& No TL
& \multicolumn{2}{|c|}{Simple TL} & \multicolumn{2}{|c}{Sophisticated TL} \\ \cline{3-7}
& &\tabincell{c}{LDA}&\tabincell{c}{CLDA}&\tabincell{c}{Imp.  (\%)} &\tabincell{c}{wAR}&\tabincell{c}{Imp. (\%)}\\ \midrule \multirow{9}{*}{1}
&CSP-LDA&      59.75$\pm$6.74&   63.27$\pm$4.37&    5.91&   64.88$\pm$4.65&    8.60\\
&CCSP-LDA&     63.54$\pm$5.52&   59.45$\pm$1.48&   -6.43&   64.95$\pm$2.18&    \underline{2.23}\\
&RCSP-LDA&     64.28$\pm$5.01&   67.60$\pm$1.29&    5.17&   70.22$\pm$1.49&    9.25\\
& EA-CSP-LDA & 59.57$\pm$6.72&   69.46$\pm$4.91&   16.61&   70.87$\pm$5.43&   18.96\\
& EA-CCSP-LDA &69.09$\pm$6.32&   79.97$\pm$0.17&   15.74&   80.37$\pm$0.30&   16.33\\
& EA-RCSP-LDA &68.24$\pm$5.31&   79.51$\pm$0.08&   16.51&   80.45$\pm$0.35&   17.90\\
& PS-CSP-LDA & 59.58$\pm$6.73&   71.51$\pm$5.26&   20.02&   71.89$\pm$5.17&   20.67\\
& PS-CCSP-LDA &67.59$\pm$5.72&   78.35$\pm$0.18&   15.92&   79.15$\pm$0.14&   17.11\\
& PS-RCSP-LDA &68.92$\pm$5.25&   79.15$\pm$0.13&   14.84&   79.56$\pm$0.17&   15.44\\ \midrule \multirow{9}{*}{2a}
&CSP-LDA&      62.54$\pm$4.88&   64.59$\pm$1.67&    3.27&   66.43$\pm$2.12&    6.22\\
&CCSP-LDA&     65.91$\pm$3.67&   68.44$\pm$0.51&    3.83&   70.32$\pm$1.08&    6.69\\
&RCSP-LDA&     65.76$\pm$3.65&   70.54$\pm$0.63&    7.26&   71.78$\pm$0.68&    9.16\\
& EA-CSP-LDA  &62.43$\pm$4.89&   70.46$\pm$1.62&   12.86&   70.76$\pm$1.76&   13.35\\
& EA-CCSP-LDA &66.97$\pm$4.28&   73.84$\pm$0.10&   10.26&   74.19$\pm$0.50&   10.77\\
& EA-RCSP-LDA &68.64$\pm$3.99&   74.76$\pm$0.55&    8.92&   75.10$\pm$0.54&    9.41\\
& PS-CSP-LDA & 62.47$\pm$4.90&   70.49$\pm$1.59&   12.85&   70.85$\pm$1.70&   13.42\\
& PS-CCSP-LDA &66.48$\pm$4.17&   73.87$\pm$0.14&   11.12&   74.67$\pm$0.47&   12.32\\
& PS-RCSP-LDA &68.02$\pm$3.89&   74.22$\pm$0.66&    9.11&   74.68$\pm$0.70&    9.78\\ \bottomrule
\end{tabular}
\end{table}

\section{Discussion} \label{sect:discussion}

The section discusses the TL pipeline in offline cross-subject MI classification using deep learning, online cross-subject classification, and offline cross-session classification.

\subsection{Offline Cross-Subject Classification Using Deep Learning}

The previous section verified the effectiveness of TL in a traditional machine learning pipeline. Deep learning has made significant breakthroughs in many fields, including EEG-based BCIs. This subsection considers the TL pipeline in offline cross-subject MI classification using deep learning models.

Two popular convolutional neural network (CNN) classifiers, EEGNet \cite{EEGNet} and ShallowCNN \cite{Schirrmeister2017}, were used in our experiments. EEGNet is a compact convolutional network with only about 1,000 parameters (the number may change slightly according to the nature of the task) for EEG-based BCIs. It introduces depthwise and separable convolutions into the construction of EEG-specific CNNs, which encapsulate well-known EEG feature extraction concepts and simultaneously reduce the number of model parameters. ShallowCNN has a very shallow architecture, consisting of a convolutional block and a classification block. The convolutional block is specially designed to handle EEG signals.

Because CNN models perform simultaneously spatial filtering, feature engineering and classification, and their computational cost is much higher than traditional machine learning models, we only compared the performances with and without EA in Figure~\ref{fig:TLBCI}. More specifically, we considered offline unsupervised cross-subject classification, i.e., all samples from the target subject were unlabeled ($N_l=0$), and all samples from the source subjects were combined and partitioned into 80\% training and 20\% validation (for early stopping). We used Adam optimizer, cross entropy loss, and batch size 32. The experiments were repeated 15 times for each target subject.

The results are shown in Table~\ref{tab:deep}. Clearly, using EA can improve the classification performance of both deep learning models, especially on Dataset~1. These results are consistent with a more comprehensive study in \cite{Kostas2020}, which shows that EA generally benefits deep learning classification of movement and motor imagery, P300, and error related negativity.

\begin{table}[htpb]
\centering \renewcommand{\arraystretch}{1.2} \setlength{\tabcolsep}{2.5mm}
\caption{The effect of EA in deep learning. Performance improvements not statistically significant ($\alpha=0.05$) are marked with an underline.} \label{tab:deep}
\begin{tabular}{c|c|ccc} \toprule
Dataset   &Model  &w/o EA (\%)&w/ EA  (\%)&Improvement (\%) \\ \midrule
\multirow{2}{*}{1}
&EEGNet    &       59.35$\pm$4.80&   70.00$\pm$8.14&    17.94 \\
&ShallowCNN&       62.85$\pm$7.47&   73.92$\pm$8.48&    17.61 \\ \midrule
\multirow{2}{*}{2a}
&EEGNet    &       72.71$\pm$3.62&   75.84$\pm$4.61&    \underline{4.30} \\
&ShallowCNN&       68.00$\pm$3.43&   73.46$\pm$1.76&    3.59 \\ \bottomrule
\end{tabular}
\end{table}

\subsection{Online Cross-Subject Classification}

All previous results considered offline cross-subject MI classification. It is also interesting to study if the TL pipeline in Figure~\ref{fig:TLBCI} can be used in online cross-subject MI classification.

To this end, we make sure EA and PS use only the available labeled target domain samples in computing the reference matrix and performing data alignment in the target domain. We also replace wAR by OwAR \cite{drwuTHMS2017}, which does not make use of offline unlabeled samples in the target domain in classifier training.

\begin{figure*}[htpb] \centering
\subfigure[]{\label{fig:MI1online}     \includegraphics[width=\linewidth,clip]{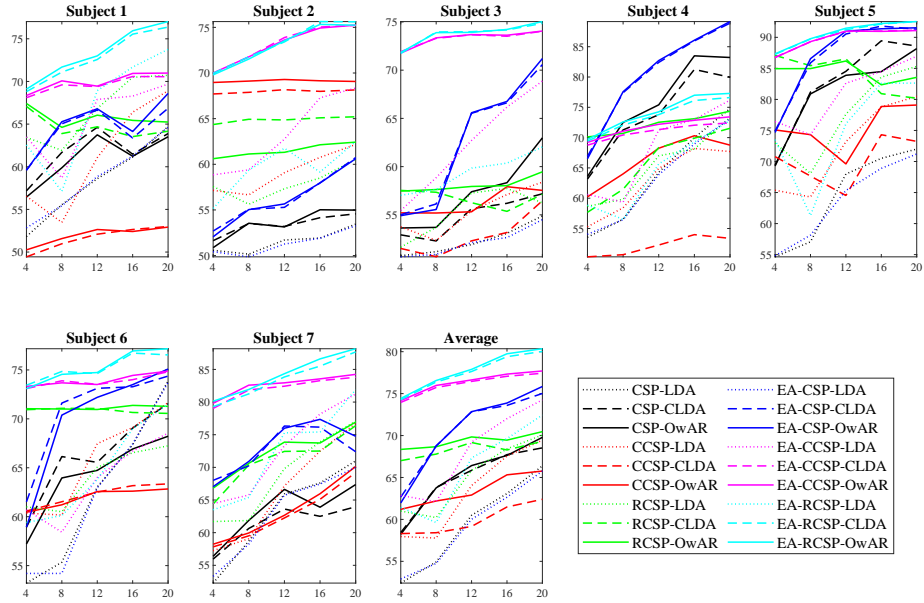}}
\subfigure[]{\label{fig:MI2online}     \includegraphics[width=\linewidth,clip]{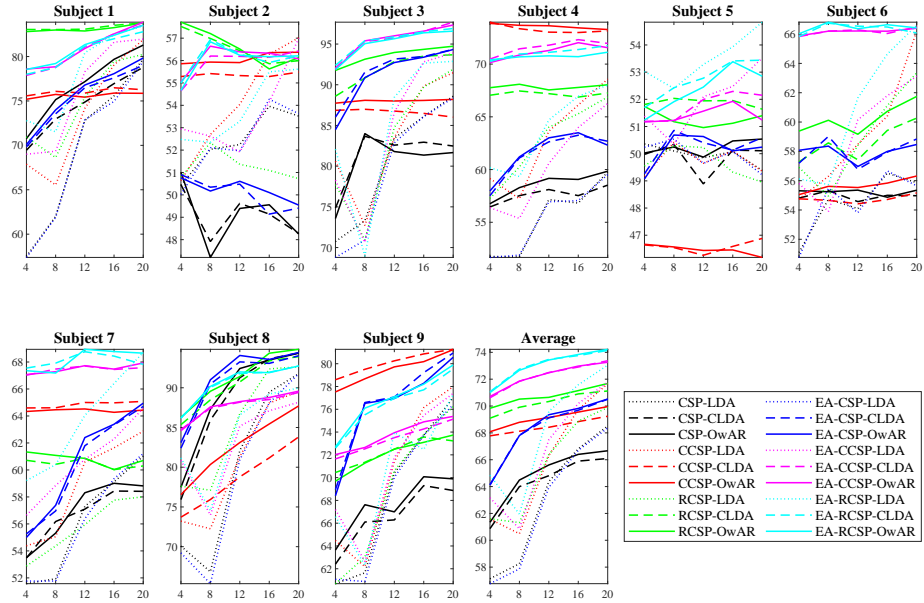}}
\caption{Online cross-subject classification accuracies (vertical axis) on the MI datasets, with different $N_l$ (horizontal axis). (a) Dataset~1; (b) Dataset~2a.} \label{fig:online}
\end{figure*}

The online cross-subject classification results are shown in Figure~\ref{fig:online} and Table~\ref{tab:7}. Generally, all observations made from offline cross-subject classification in the previous section, e.g., considering TL in more components of Figure~\ref{fig:TLBCI} benefits the performance more, and data alignment is very important to subsequent TL, still hold in online cross-subject classification.

Comparing Tables~\ref{tab:1} and \ref{tab:7} shows that the offline classification performances were generally slightly better than their online counterparts, which is intuitive, as offline classification makes use of the unlabeled target domain samples, which provides extra information in EA, PS and wAR.

\begin{table}[htpb]
\centering \renewcommand{\arraystretch}{1.2} \setlength{\tabcolsep}{1.5mm}
\caption{Online cross-subject classification accuracies (mean$\pm$std) of different TL approaches, and their improvements over CSP-LDA. Performance improvements not statistically significant ($\alpha=0.05$) are marked with an underline.}
 \label{tab:7}
\begin{tabular}{c|cc|cc} \toprule
\multirow{3}{*}{Algorithm}&\multicolumn{2}{c|}{Dataset 1} &\multicolumn{2}{|c}{Dataset 2a}\\ \cline{2-5}
& Accuracy & Improvement & Accuracy & Improvement \\
& (\%) & (\%) & (\%) & (\%) \\ \midrule
CSP-LDA &      59.46$\pm$5.75&      --&   63.02$\pm$5.09&      --   \\
CSP-CLDA&      64.87$\pm$4.03&    9.10&   64.33$\pm$2.11&    \underline{2.08}   \\
CSP-OwAR&      65.16$\pm$4.45&    9.59&   64.90$\pm$2.16&    2.98   \\
CCSP-LDA&      63.35$\pm$5.37&    6.55&   65.96$\pm$4.90&    4.66   \\
CCSP-CLDA&     59.95$\pm$1.87&    \underline{0.83}&   68.48$\pm$0.58&    8.66   \\
CCSP-OwAR&     63.48$\pm$1.98&    6.76&   69.10$\pm$0.72&    9.66   \\
RCSP-LDA&      65.11$\pm$4.38&    9.50&   65.61$\pm$4.00&    4.12   \\
RCSP-CLDA&     68.31$\pm$0.96&   14.89&   70.27$\pm$0.81&   11.51   \\
RCSP-OwAR&     69.35$\pm$0.86&   16.64&   70.76$\pm$0.69&   12.29   \\ \midrule
EA-CSP-LDA&    59.22$\pm$5.37&   \underline{-0.41}&   62.73$\pm$5.20&   -0.47   \\
EA-CSP-CLDA&   70.58$\pm$5.00&   18.70&   68.26$\pm$2.49&    8.31   \\
EA-CSP-OwAR&   70.66$\pm$5.51&   18.83&   68.32$\pm$2.56&    8.41   \\
EA-CCSP-LDA&   68.09$\pm$5.43&   14.52&   66.69$\pm$4.61&    5.83   \\
EA-CCSP-CLDA&  76.13$\pm$1.38&   28.03&   72.25$\pm$1.07&   14.64   \\
EA-CCSP-OwAR&  76.36$\pm$1.39&   28.43&   72.26$\pm$1.02&   14.66   \\
EA-RCSP-LDA&   66.10$\pm$5.43&   11.16&   68.20$\pm$4.67&    8.23   \\
EA-RCSP-CLDA&  77.55$\pm$2.33&   30.42&   73.04$\pm$1.22&   15.90   \\
EA-RCSP-OwAR&  77.80$\pm$2.41&   30.84&   73.03$\pm$1.27&   15.89   \\ \midrule
PS-CSP-LDA&    59.23$\pm$5.53&   -0.38&   62.75$\pm$5.23&   -0.43   \\
PS-CSP-CLDA&   71.45$\pm$4.84&   20.17&   69.04$\pm$2.32&    9.55   \\
PS-CSP-OwAR&   71.26$\pm$5.33&   19.84&   68.96$\pm$2.48&    9.43   \\
PS-CCSP-LDA&   67.14$\pm$5.22&   12.91&   67.04$\pm$3.66&    6.39   \\
PS-CCSP-CLDA&  76.25$\pm$1.13&   28.23&   72.47$\pm$1.10&   14.99   \\
PS-CCSP-OwAR&  76.73$\pm$1.16&   29.05&   72.54$\pm$1.15&   15.10   \\
PS-RCSP-LDA&   67.60$\pm$5.10&   13.69&   67.40$\pm$4.15&    6.95   \\
PS-RCSP-CLDA&  76.62$\pm$1.16&   28.86&   72.69$\pm$1.33&   15.34   \\
PS-RCSP-OwAR&  77.02$\pm$1.07&   29.54&   72.75$\pm$1.30&   15.44   \\ \bottomrule
\end{tabular}
\end{table}

\subsection{Offline Cross-Session Classification}

It is well-known that EEG signals are non-stationary \cite{Liyanage2013}, i.e., EEG responses to the same stimulus from the same subject in different sessions are usually varying. This subsection evaluates how the proposed TL pipeline can be used to handle EEG non-stationarity in cross-session classification.

In Dataset~2a, each of the nine subjects had two sessions (training and evaluation), collected on different days. For each subject, we used the training session as the source domain, and the evaluation session as the target domain. Other experimental settings were identical to those in previous subsections, except that we used $w_t=2$ in wAR, as in cross-session TL the number of labeled source domain samples was much smaller than that in cross-subject TL.

\begin{figure}[htpb] \centering
\includegraphics[width=\linewidth,clip]{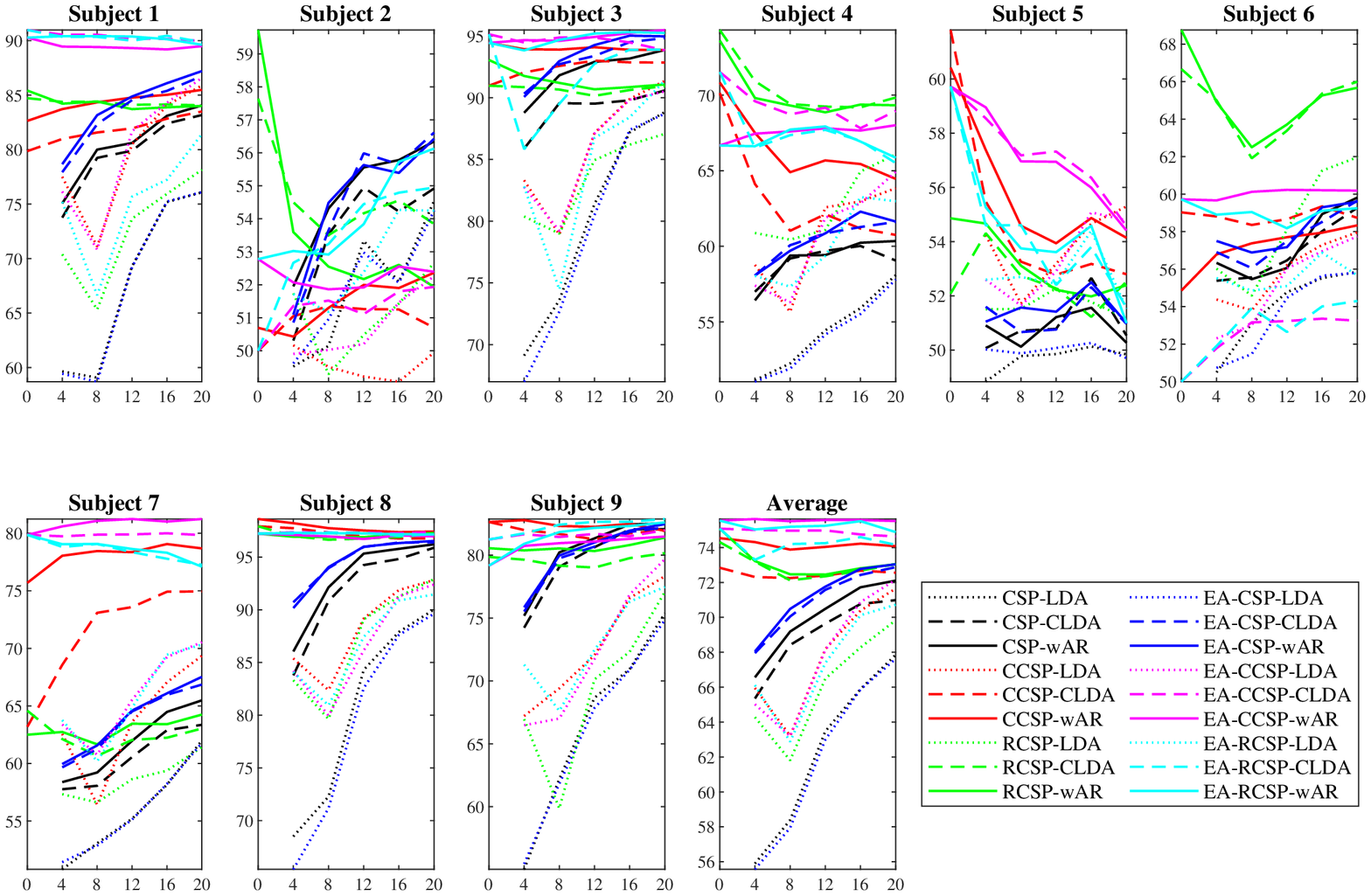}
\caption{Offline cross-session classification accuracies (vertical axis) on Dataset~2a, with different $N_l$ (horizontal axis).} \label{fig:MI2CC}
\end{figure}

The results are shown in Figure~\ref{fig:MI2CC}. Some subjects, e.g., Subjects 1, 3, 7, 8 and 9, demonstrated very stable classification performance when $N_l$ increased, indicating that their EEG signals were quite stationary, at least in the two experimental sessions. However, the remaining four subjects's EEG signals were more non-stationary, and hence the classification performance had large variations.

The average classification results are shown in Table~\ref{tab:8}. EA-CCSP-wAR, which integrates data alignment and TL in two components (spatial filtering and classification), achieved the best average performance. EA-RCSP-wAR, which was the best performer in cross-subject transfers in the previous section, was slightly worse, but still better than the other 16 approaches with or without TL. On average, almost all approaches with EA outperformed their counterparts without EA, suggesting again the importance and necessity of explicitly adding a data alignment block before TL.

\begin{table}[htpb]
\centering \renewcommand{\arraystretch}{1.2} \setlength{\tabcolsep}{3mm}
\caption{Offline cross-session classification accuracies (mean$\pm$std) of different TL approaches, and their improvements over CSP-LDA. All performance improvements were statistically significant ($\alpha=0.05$).}      \label{tab:8}
\begin{tabular}{c|cc} \toprule
Algorithm & Accuracy (\%)& Improvement (\%)\\ \midrule
CSP-LDA &    62.30$\pm$5.03&    --  \\
CSP-CLDA&    69.03$\pm$2.30&   10.81  \\
CSP-wAR&     70.01$\pm$2.24&   12.38  \\
CCSP-LDA&    67.87$\pm$3.39&    8.95  \\
CCSP-CLDA&   72.44$\pm$0.18&   16.27   \\
CCSP-wAR&    74.10$\pm$0.17&   18.95  \\
RCSP-LDA&    66.14$\pm$3.18&    6.16  \\
RCSP-CLDA&   72.68$\pm$0.45&   16.66  \\
RCSP-wAR&    72.80$\pm$0.34&   16.85  \\ \midrule
EA-CSP-LDA&  62.00$\pm$5.13&   -0.48  \\
EA-CSP-CLDA& 70.98$\pm$2.00&   13.93  \\
EA-CSP-wAR&  71.22$\pm$2.04&   14.32  \\
EA-CCSP-LDA& 67.88$\pm$3.79&    8.95  \\
EA-CCSP-CLDA&74.87$\pm$0.17&   20.17  \\
EA-CCSP-wAR& 75.56$\pm$0.05&   21.29  \\
EA-RCSP-LDA& 67.44$\pm$3.18&    8.26  \\
EA-RCSP-CLDA&74.10$\pm$0.48&   18.94  \\
EA-RCSP-wAR& 75.17$\pm$0.24&   20.66  \\ 
\bottomrule
\end{tabular}
\end{table}

In summary, TL is also effective in handling non-stationarity of EEG signals in cross-session MI classification, and considering TL in more components of the classification pipeline is generally more beneficial.

\section{Conclusions and Future Research} \label{sect:conclusions}

Transfer learning has been widely used in MI-based BCIs to reduce the calibration effort for a new subject, and demonstrated promising performance. While a closed-loop MI-based BCI system, after EEG signal acquisition and temporal filtering, includes spatial filtering, feature engineering, and classification blocks before sending out the control signal to an external device, previous approaches only considered TL in one or two such components.

This paper proposes that TL could be considered in all three components, and it is also very important to specifically add a data alignment component before spatial filtering to make the source domain and target domain data more consistent. Offline and online classification experiments on two MI datasets verified that:
\begin{enumerate}
\item Generally, using TL in different components of Figure~\ref{fig:TLBCI} can achieve better classification performance than not using it, in both cross-subject and cross-session classification, for both online and offline classification.
\item Generally, a more sophisticated TL approach outperforms a simple one.
\item Data alignment is a very important preprocessing step in TL. It benefits both traditional machine learning and deep learning.
\item TL in different components of Figure~\ref{fig:TLBCI} could be complementary to each other, so integrating them can further improve the classification performance.
\end{enumerate}

The following directions will be considered in our future research:
\begin{enumerate}
\item Compared with other components, not enough attention has been paid to TL in feature engineering of BCI systems. We will develop more sophisticated TL approaches for feature engineering in the future, and also other components in Figure~\ref{fig:TLBCI}.

\item This paper considers only binary MI classification problems in BCIs. We will extend the analysis to other BCI classification paradigms, e.g., event-related potentials and steady-state visual evoked potentials, and also BCI regression problems, e.g., driver drowsiness estimation \cite{drwuFWET2019,drwuTFS2017} and user reaction-time estimation \cite{drwuRG2017}. Furthermore, we will also consider TL in multi-class classification.

\item It has been shown \cite{drwuSMC2014,drwuTNSRE2016} that integrating TL with active learning \cite{drwuRSVP2016} in the classifier can further improve the offline classification performance. It is interesting to study if TL and active learning can be integrated in other components of the BCI system, e.g., spatial filtering and feature engineering.
\end{enumerate}

\end{document}